\documentstyle[spie]{article} 
\input{plot.sty}   

\title{Keck studies of M31's stellar halo\footnote{~Based in part on data
collected at W.$\,$M.\ Keck Observatory, which is operated as a scientific
partnership among the California Institute of Technology, the University of
California and the National Aeronautics and Space Administration.  The
Observatory was made possible by the generous financial support of the
W.$\,$M.\ Keck Foundation.}}

\author{Puragra Guhathakurta\supit{a}, David B.\ Reitzel\supit{a}, and
Eva K.\ Grebel\supit{b} 
\skiplinehalf 
\supit{a}UCO/Lick Obs, Univ of California at Santa Cruz, 1156 High St,
\\ Santa Cruz, California 95064, USA\\
\skiplinehalf 
\supit{b}Univ of Washington, Dept of Astronomy, Box 35158,
\\ Seattle, Washington~98195-1580, USA
}

\authorinfo{Further author information: (Send correspondence to P.G.)\\
P.G.:  Alfred P.\ Sloan Research Fellow;  E-mail: raja@ucolick.org\\
D.B.R.:  E-mail: reitzel@ucolick.org\\
E.K.G.:  Hubble Fellow;  E-mail: grebel@astro.washington.edu}

\begin{document} 
  \maketitle 

\begin{abstract}
We present Keck 10-meter / LRIS spectra of candidate red giants in the halo
of M31, located at a projected radius of $R=19$~kpc on the minor axis.  These
spectroscopic targets have been selected using a combination of $UBRI$-based
and morphological screening to eliminate background galaxies.  Radial
velocity measurements are used to separate M31 halo giants from foreground
Milky Way dwarf stars, M31 disk stars, and residual background galaxies.  The
metallicity of each M31 halo giant is measured using standard photometric and
spectroscopic techniques, the latter based on the strength of the Ca~II
triplet.  The various [Fe/H] estimates are in rough agreement with one
another.  The data reveal a large spread ($>2$~dex) in [Fe/H] in M31's halo;
there is no strong radial [Fe/H] gradient.  LRIS and HIRES spectra are also
presented for red giants in five dwarf spheroidal satellites of M31: And\,I,
And\,III, And\,V, And\,VI, and And\,VII.  There appears to be a significant
metallicity spread in And\,VI and possibly in And\,I.  The new radial
velocity data on these outer dwarfs are used to constrain the total mass of
M31: the best estimate is under $10^{12}\,M_\odot$, somewhat less than the
best estimate for the Milky Way.
\end{abstract}



\section{OVERVIEW}
\label{sect:overview}  

This paper describes two ongoing observational programs at Keck unified by a
common goal: to investigate the metallicity, structure, and dynamics of M31's
extended stellar halo.  The first is a low-resolution spectroscopic survey of
field red giant stars in M31's outer halo ($R=10\>$--$\>$50~kpc), while the
second aims to combine low-, intermediate-, and high-resolution spectroscopy
of giants in dwarf spheroidal (dSph) companions of M31 located at
$R=50\>$--$\>$300~kpc.  Such dwarfs have long been thought of as basic
building blocks in the process of galaxy formation.  The metallicity
distribution of our target stars serves as as a fossil record of the
formation of M31's halo.  Moreover, these stars are good tracers for
investigating the internal dynamics of the dwarfs as well as the global
dynamics of the M31 subgroup, and hence the dark matter distribution within
the Local Group.

\section{M31 FIELD HALO RED GIANTS}
\label{sect:m31halo}

\subsection{Introduction: Motivation and Background}
\label{sect:2.1}

The radial abundance gradient and abundance spread in M31's stellar halo can
potentially discriminate
between competing galaxy formation models.  The dissipational collapse
model\cite{eggen,larson} predicts a strong radial
metallicity gradient, whereas the accretion model\cite{searle}
predicts no strong gradient.
It is important to study both globular cluster and field star populations
since they may be dynamically distinct from each other.
The Andromeda galaxy (M31) provides a global external perspective 
of a large spiral much like our own, and yet is close enough for individual
stars to be resolved.  Ref.~\citenum{huchra} shows evidence
for a weak metallicity gradient in a sample of 150~M31 globular clusters,
with a mean iron abundance of $\rm[Fe/H]=-1.2$, which is slightly higher than
the mean value of $\rm[Fe/H]=-1.4$ for Galactic globular clusters.

The metallicity of M31 field halo stars remains uncertain even after a decade
or more of concerted effort using ground-based telescopes.
Ref.~\citenum{mould} found a metallicity of $\rm[Fe/H]\approx-0.6$ for stars
located at a projected distance of $R=7$~kpc along the minor axis and
Ref.~\citenum{pritchet88} estimated $\rm[Fe/H]\approx-1.0$ at $R=8.6$~kpc
also on the minor axis.  A study of an $R=16$~kpc field around the globular
cluster G219 suggested a large metallicity spread\cite{christian}, with a
mean value $\rm[Fe/H]\geq-1.0$, as did the studies described in
Ref.~\citenum{durrell}, Ref.~\citenum{davidge} ($R=6.7$~kpc),
and Ref.~\citenum{couture} which targeted fields around five~M31 globular
clusters.  Recent {\it Hubble Space Telescope\/} ({\it HST\/}) studies have
used its excellent angular resolution to separate field stars from background
galaxies and find the mean metallicity at various points in the M31 spheroid
to be comparable to that of 47~Tuc: $\rm[Fe/H]\sim-0.7$\cite{holland,rich}.

The ground-based studies are hindered to one degree or another by sample
contamination by distant background field galaxies many of which are
compact enough to be mistaken for stars at faint magnitudes.  The {\it HST\/}
studies, on the other hand, are limited by a small field of view and the lack
of comparison field control samples.  Studies of the inner halo risk possible
contamination by M31 disk stars, while those of the sparse outer halo have
to contend with a significant fraction of foreground contaminants, Milky Way
dwarf stars at M31's low Galactic latitude.  In general, sample contamination
makes it difficult to draw conclusions about the average [Fe/H] of M31's
stellar halo and especially its [Fe/H] spread.

\subsection{Photometric and Morphological Screening for Red Giants in M31}
\label{sect:2.2}

Our study is based on deep $UBRI$ images, obtained with the KPNO 4-meter
telescope, of a $16'\times16'$ field at $R=19$~kpc
($1.6^\circ$) on M31's SE minor axis.  This field is further out on the minor
axis than the ones used in previous studies, probes at least as far out into
the M31 spheroid as the others (based on a 2:1 flattened spheroid), and is
the least susceptible to contamination by the M31 disk.  Observations of a
well-matched comparison field, well away from M31 but at a comparable Galactic
latitude, serve as a control data set.  Four high resolution $I$-band
images obtained with Keck, covering most of the field, provide morphological
information.

The stars in these fields (M31 giants and Milky Way dwarfs) are outnumbered
by the dense background of distant faint blue field galaxies, many of which
are too compact to be distinguished from stars on the basis of image
morphology alone (even with Keck's $0.7''$--$0.8''$
seeing).  We use an alternative method developed in Ref.~\citenum{gould}
for discriminating between stars and faint galaxies, one that uses broadband
$UBRI$ color information in addition to morphological information.  The
technique relies on the fact that stars, even those spanning a wide range of
age and metallicity, occupy a narrow, well-defined locus in multicolor
space whereas galaxies display a broad distribution due to their mix of
stellar populations and redshifts.  For an object to be identified as a
stellar candidate, its intensity profile must be consistent with the PSF
{\it and\/} its $U-B$, $B-R$, and $R-I$
colors must be consistent (within errors) with the stellar locus defined by
model isochrones\cite{bertelli} and empirical data on Galactic
globular cluster giants.

This morphological and $UBRI$ color-based screening is done on all objects
detected in the M31 halo and comparison fields.  The color-magnitude diagram
(CMD) of surviving objects in the comparison field is well described by a
superposition of foreground Galactic dwarf stars (in keeping with a standard
empirical model of the Galaxy) against a backdrop of residual contaminating
faint blue field galaxies, while the M31 halo field contains a clear excess
of faint red objects ($I\sim20\>$--$\>$23, $B-I\sim2\>$--$\>$3.5) in 
addition to these two components.  The location of this population of faint
red objects in the CMD is consistent with what one would expect for red
giant stars at the distance of M31 (Fig.~1).  We have carried out a
statistical subtraction between the M31 halo field and comparison field
CMDs.

A detailed description of the selection process and results is given in
Ref.~\citenum{reitzel98}.

\begin{figure}
\plotfiddle{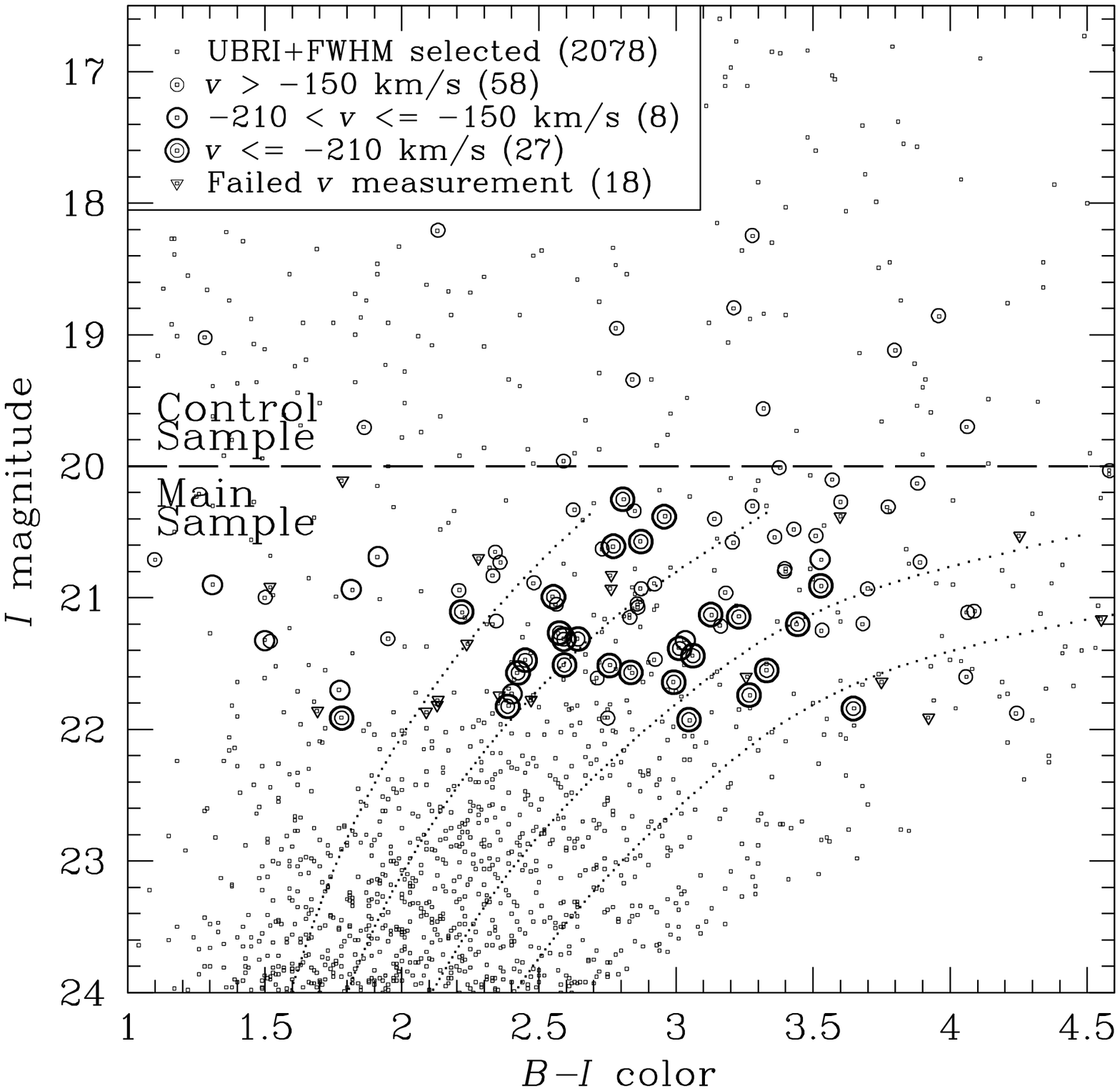}{1.6truein}{0}{38}{38}{-249}{-137}
\plotfiddle{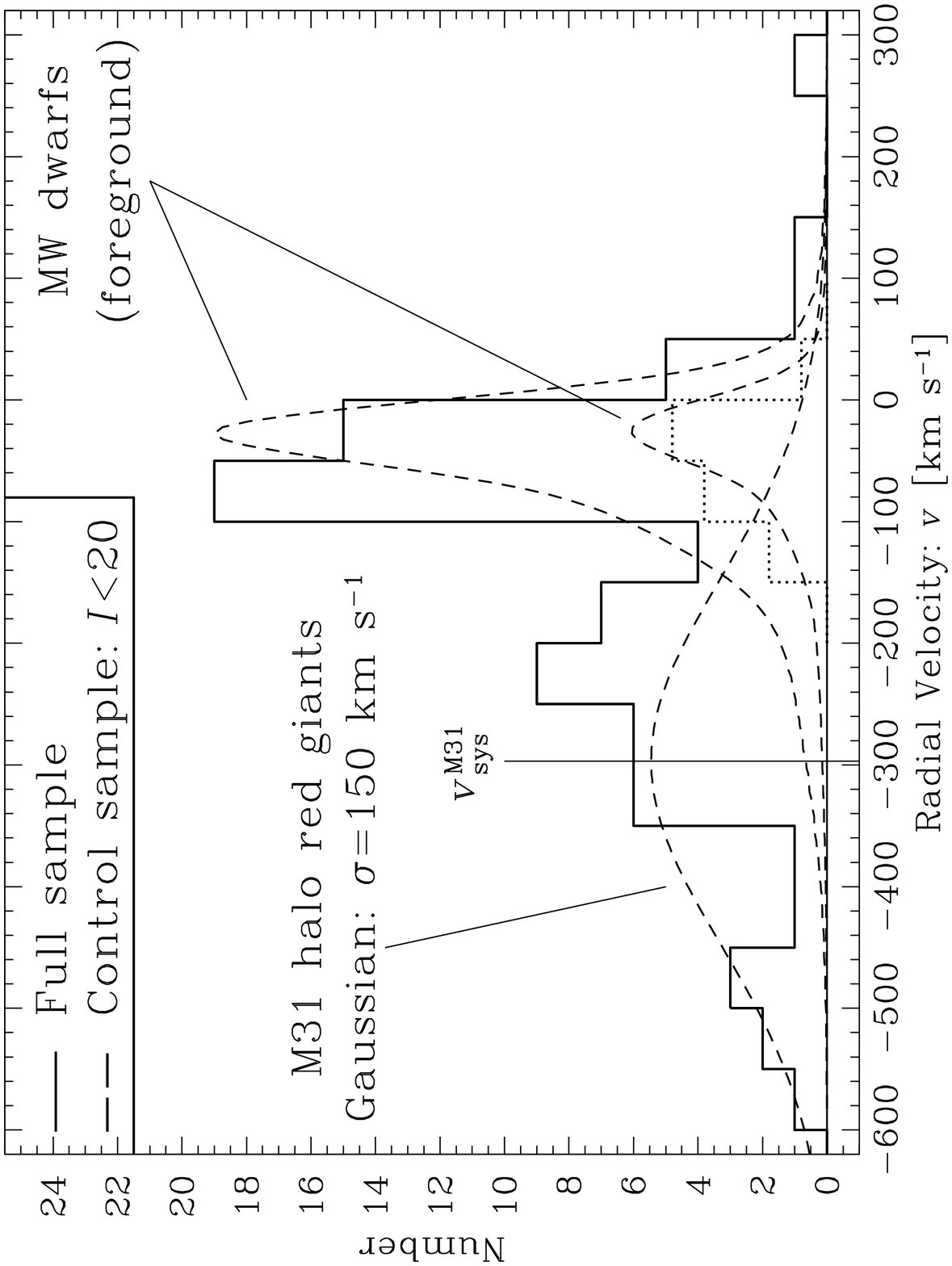}{1.6truein}{-90}{38}{38}{-40}{273}
\end{figure}
\begin{figure}
\vskip -2.25truecm
\begin{minipage}{8cm}
\caption[m31halo_cmd]{ \label{fig:m31halo_cmd}
Color-magnitude plot of objects in the M31 halo field whose $UBRI$ colors and
angular sizes are consistent with stars.  Keck spectroscopy targets are
plotted with special symbols\cite{reitzel00}.  The control sample consists of
objects with $I<I_{\rm TRGB}(\rm M31)$.  The model isochrones are from
Ref.~\citenum{vandenberg} for (L$\rightarrow$R): $\rm[Fe/H]=-2.31$, $-1.41$,
$-0.71$, and $-0.30$.}

\end{minipage}
\hfill
\begin{minipage}{9cm}
\caption[m31halo_vel]{ \label{fig:m31halo_vel}
Distribution of radial velocities derived from Keck spectra of 99~target red
giant candidates (solid histogram) and 12~control sample objects (dotted
histogram) selected on the basis of $UBRI$ colors and
morphology\cite{reitzel00}.  Two model curves from
Ref.~\citenum{ratnatunga} for foreground Galactic dwarfs are shown on the
right, while the broad curve to the left is a Gaussian representing M31 halo
red giants.}
\end{minipage}
\end{figure}

\subsection{Keck Spectroscopy}
\label{sect:2.3}

Of the $UBRI$- and morphology-selected sample of objects in our $R=19$~kpc
minor axis field, the brightest ones ($20<I<22$) serve as excellent targets
for follow-up spectroscopy.  Spectra of 99~M31 halo red giant candidates have
been obtained with the 10-meter Keck telescope and Low-Resolution Imaging
Spectrometer\cite{oke} (LRIS) around the Ca~II triplet
($\sim8500$--8700\,\AA).  In addition, spectra have been obtained of
12~control objects which are known to be significantly brighter than the tip
of M31's red giant branch, and which are thus likely to be foreground
Galactic dwarf stars.  Five multislit masks were observed for $\sim4$~hr
each during 4~nights in the Fall 1996 and 1997 observing seasons, with
$\sim20\>$--$\>$25~objects per mask.  The observations used a
1200~lines~mm$^{-1}$ grating and $1''$-wide slitlets, for a dispersion of
0.62\,\AA~pixel$^{-1}$ and resolution of $\sigma\sim40$~km~s$^{-1}$ over the
spectral range of 7500--9000$\,$\AA.

The radial velocity of each target object is determined from its spectrum
using standard cross-correlation techniques.  This kinematic information is
useful for distinguishing among M31 halo giants, M31 disk giants, foreground
Galactic dwarfs, and distant background galaxies.
Most of the foreground contamination in this low Galactic latitude M31 field
($\vert{b}\vert=22^\circ$) is expected to be due to dwarf stars in the Milky
Way's disk with heliocentric velocities close to zero, significantly
displaced from M31's systemic velocity of $-297$~km~s$^{-1}$.  M31 halo red
giants are expected to have a large spread of radial velocities due to
their random motion in the galaxy's potential, while M31 disk giants are
expected to cluster tightly around its systemic velocity (the line-of-sight
component of the disk rotation velocity is expected to be zero as our field
is on the minor axis).

In addition to the removal of residual contaminants, spectroscopy allows
direct measurement of the metallicity of each star using the near infrared
Ca~II absorption line strengths, $\rm\sum{Ca}$ (sum of the three equivalent
widths).  There has been a lot of work to develop calibration relations based
on observations of the line strengths of red giants in fiducial Galactic star
clusters with known
metallicities\cite{olszewski,armand91,rutledge97a,rutledge97b}.
Such relations connect $\rm\sum{Ca}$ and the brightness of the star
$\Delta{V}$ relative to the horizontal branch to [Fe/H], and have
been calibrated over the range $\rm-2.5\leq[Fe/H]\leq0.0$.  The line strength,
$\rm\sum{Ca}$, of each of our M31 halo red giant candidates is determined
using an empirical linear relation between the relative strength of the cross
correlation peak and the directly-measured, but somewhat noisier statistic
$\rm\sum{Ca}$ for selected stars.  This derived $\rm\sum{Ca}$ is then
used to calculate the metallicity of each object.  A detailed description is
given in Ref.~\citenum{reitzel00}.

\subsection{Results}
\label{sect:2.4}

\subsubsection{Density of M31's stellar halo}
\label{sect:2.4.1}

The surface density of red giant candidates in the $R=19$~kpc M31 halo field,
as derived from the statistically-subtracted, $UBRI$ color- and
morphologically-selected sample, is coupled
with the {\it HST\/} studies in Ref.~\citenum{holland}.
The data indicate that M31's stellar halo is much denser
and/or larger than that of the Galaxy: $(\rho_{\rm M31}^{\rm RGB}/\rho_{\rm
MW}^{\rm RGB})(\Lambda/1.5)^{-\nu}\sim10$, where $\Lambda$ is the
ratio of the radial scale lengths of M31 and the Galaxy and $\nu=-3.8$
is the assumed power law index of the density profile\cite{reitzel98}.  The
density of M31's inner halo may be even higher than this 
estimate: while we have adopted a single power law profile for M31's halo,
the outer profile is known to steepen both from the counts in our
$R=19$~kpc field and from Ref.~\citenum{pritchet94} which finds
that the profile is better fit by a de~Vaucouleurs law than a power law.
A refined estimate of the halo density profile and flattening are underway
based on a spectroscopically-selected sample of an additional 50--100~secure
M31 halo red giants at various locations on the major and minor axes.  Radial
velocities and line strengths derived from this sample should enhance our
knowledge of the dynamics and metallicity of M31's halo, respectively (see
Sec.~2.4.2 and 2.4.3 below).

\begin{figure}
\plotfiddle{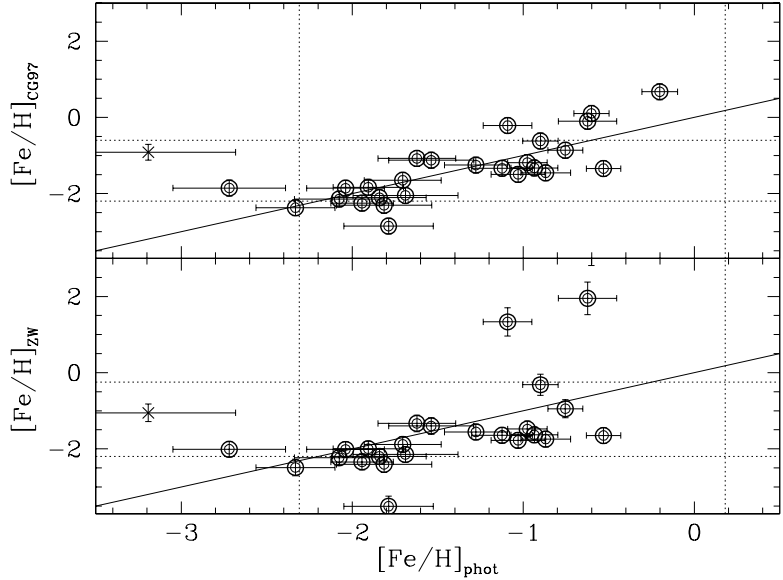}{2.7472truein}{-90}{40}{40}{-170}{229}
\caption[m31halo_fehcomp]{ \label{fig:m31halo_fehcomp}
A star-by-star comparison of the metallicities derived using various methods:
photometric versus spectroscopic (both CG97 and ZW calibrations from
Ref.~\citenum{rutledge97b}).  Only the secure sample of M31 halo red giants
with $v<-210$~km~s$^{-1}$ is shown.  The horizontal and vertical dashed lines
indicate the range of [Fe/H] over which each method is calibrated.  The
photometric estimate of [Fe/H] agrees better with the CG97 calibration of the
Ca~II absorption line strength than with the ZW calibration.}
\end{figure}

\subsubsection{M31 halo kinematics and residual sample contaminants}
\label{sect:2.4.2}

The radial velocities measured from the $>100$ Keck spectra presented in this
paper display a bimodal distribution with one component centered on M31's
systemic velocity ($v_{\rm sys}=-297$~km~s$^{-1}$) and another component
centered at $\approx0$~km~s$^{-1}$ (Fig.~2).  We use an empirical star count
model of the Galaxy (IASG model)\cite{ratnatunga} to estimate the expected
distribution of line-of-sight velocities of foreground dwarfs.  The
spectroscopic control sample consists of 12~stars with $I<20$, which is
brighter than $I_{\rm TRGB}$ at the distance of M31 (Fig.~1); these stars are
thus exclusively foreground dwarfs.  The IASG model matches their velocity
distribution very well (dotted histogram in Fig.~2).

A linear combination of the IASG model $v$ distribution and a Gaussian
centered on M31's systemic velocity is fit to the data; the best fit Gaussian
width is $\sigma=150_{-30}^{+50}$~km~s$^{-1}$ and the fraction of M31 halo
stars in the sample of 99 targets is estimated to be $\sim50\%$.  Note, with
this normalization of the Galactic dwarf $v$ distribution (see Fig.~2),
almost all the stars with $v>-50$~km~s$^{-1}$ are likely to be foreground
dwarfs. The model results may be used to estimate the foreground contamination
fraction for various velocity cuts in the sample---e.g.,~$v>-150$~km~s$^{-1}$
corresponds to $>$50\% contamination in all velocity bins, while
$v<-210$~km~s$^{-1}$ corresponds to $<$25\% contamination.  The different
symbols in the CMD plot (Fig.~1) are based on these velocity ranges.  The
lack of a significant concentration of objects at the
exact systemic velocity indicates a negligibly small degree of contamination
by M31 disk giants.  Selecting stars with $v<-210$~km~s$^{-1}$ yields a secure
sample of M31 halo red giant stars.

\begin{figure}
\plotfiddle{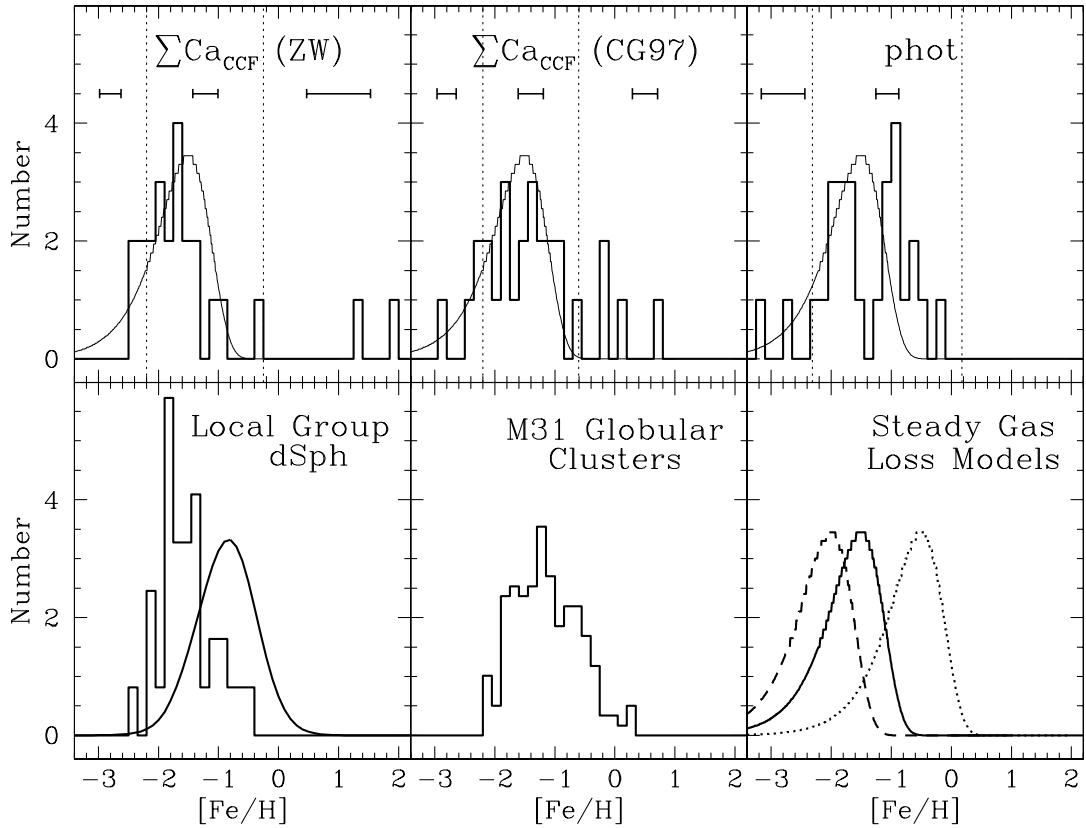}{1.56truein}{-90}{59}{59}{-240}{153}
\vskip 6.3truecm
\caption[m31halo_feh]{ \label{fig:m31halo_feh}
The metallicity distribution of spectroscopically-selected secure M31 halo
red giant stars.  From left to right, the upper panels show [Fe/H] estimates
derived from the Ca~II line strength with ZW and CG97 calibration
relations\cite{rutledge97b}, and from the ($I$, $B-I$) CMD.  The histogram
and curve in the lower left panel show unweighted and luminosity-weighted
[Fe/H] distributions of Local Group dwarf galaxies, respectively.  The
weighted distribution takes the [Fe/H] spread within each dwarf into
account; it is dominated by a few dwarfs (LMC, SMC, M32, NGC~205, etc.).  The
lower middle panel shows the unweighted distribution of M31 globular
clusters.  The lower right panel shows chemical enrichment models with steady
gas loss\cite{reitzel00}.}
\end{figure}

\subsubsection{Radial gradient and spread in metallicity}
\label{sect:2.4.3}

The secure sample of M31 halo red giant stars with $v<-210$~km~s$^{-1}$ is
particularly useful for investigating the mean [Fe/H] and spread.  The iron
abundance of each of these stars is measured using two independent methods:
the star's location in the $B-I$ vs.\ $I$ CMD and the strength of the Ca~II
absorption lines, $\rm\sum{Ca}$.  The CMD-based photometric method compares
each star to model isochrones calculated over a range of
metallicities\cite{bertelli,vandenberg}.  The line strength $\rm\sum{Ca}$ is
calibrated to [Fe/H] using two independent calibration
relations\cite{zinn,carretta}, labeled ``ZW'' and ``CG97'', respectively,
following the designation in Ref.~\citenum{rutledge97b}.  The photometric
estimate of [Fe/H] is in reasonable agreement with the spectroscopic
estimates within the calibrated range of each method, especially for the CG97
calibration (Fig.~3).

Measuring the abundance gradient versus radius in M31's halo involves
comparison of the mean [Fe/H] in our data set to {\it HST}-based photometric
determinations of [Fe/H] at $R=7$ and 11~kpc on the minor axis\cite{holland}
and at $R=40$~kpc on the major axis\cite{rich}.  Both of our $R=19$~kpc
samples of halo red giants, the large $UBRI$- and morphologically-selected
statistically-subtracted sample and the smaller but cleaner
spectroscopically-selected secure sample, yield $\rm\langle[Fe/H]\rangle$ in
the vicinity of $-1.6$ to $-1.3$ using various ways to estimate [Fe/H]
(photometric and spectroscopic).  This mean abundance is roughly consistent
with the results of the {\it HST\/} studies indicating that M31 lacks a
strong radial gradient in metallicity.  However, the use of heterogeneous
samples, different [Fe/H] determination methods, and possible sample
contamination (M31 disk stars) in the inner halo fields, makes this
conclusion uncertain.  We are in the process of isolating clean samples of
halo red giants at a variety of radii in M31 and determining [Fe/H] in a
uniform way with the help of Keck LRIS spectra (Sec.~2.4.1).

\begin{figure}
\plotfiddle{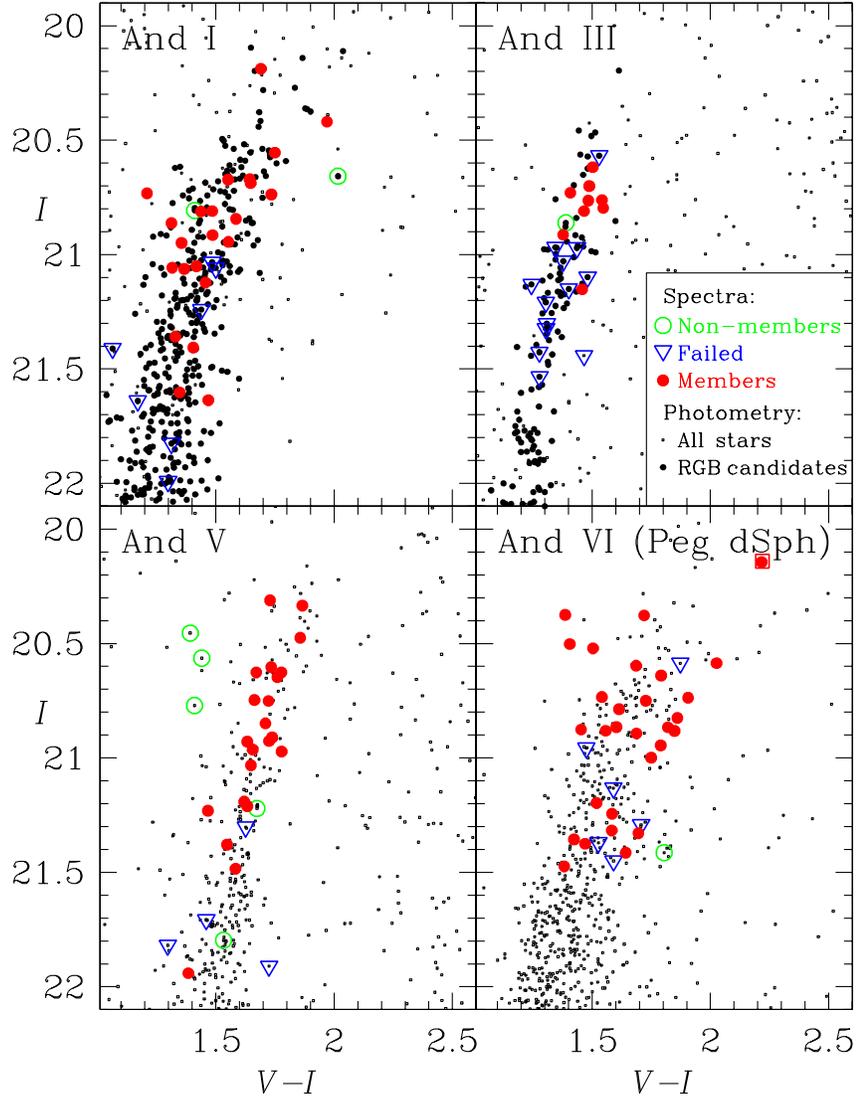}{5.462truein}{0}{60}{60}{-193}{-39}
\caption[m31dsph_cmd]{ \label{fig:m31dsph_cmd}
Color-magnitude diagrams for four dSph companions of M31 zoomed in around the
tip of the red giant branch showing the full photometric data sets (smallest
black dots) from which the LRIS spectroscopic targets are drawn.
The slightly bolder black dots in the two upper panels represent giant
candidates identified on the basis of Washington DDO51 photometry.
Spectroscopic targets can be divided into: dSph
members (filled red circles), field/foreground interlopers (open green
circles), and failed velocity measurements (open blue triangles).  The
brightest of the And\,VI member stars (filled red circle with square) was
included as a mask alignment star; it is slightly brighter
than the giant branch tip.  See Ref.~\citenum{ggprpvom} for details.}
\end{figure}

The secure sample of M31 halo red giants in the $R=19$~kpc field displays a
large spread in [Fe/H] ($\sim2$~dex), independent of the metallicity
measurement method (upper panels of Fig.~4).  The metallicity distribution
among M31 field halo stars is similar to the distribution of mean [Fe/H]
values of M31 globular clusters and Local Group dwarf satellite galaxies
(Fig.~4).  The observed [Fe/H] distribution in M31's halo can be explained in
terms of chemical enrichment models with steady gas loss from the system (see
details in Ref.~\citenum{reitzel00}).
 
\section{M31'S DWARF SPHEROIDAL SATELLITES}
\label{sect:m31dsph}

\subsection{Introduction: Basic Building Blocks of Galaxy Formation}
\label{sect:3.1}

Dwarf galaxies, the most numerous type of galaxy, are considered to be major
building blocks for the formation and evolution of more massive
galaxies\cite{grebel97,mateo,grebel99}.  Among dwarf galaxies, the least
luminous, least massive morphological type known is the dwarf spheroidal
subclass.  Almost all known dSphs are found in close proximity to more
massive galaxies, and are therefore particularly useful as tracers of the
gravitational potential of the ``parent'' galaxy as well as for the study
of environmental effects on dwarf galaxy evolution.

While the nine dSph companions of the Milky Way have been studied in great
detail, similar studies of the M31 dSphs have only recently become feasible
thanks to large-aperture 8--10-meter class telescopes and efficient
modern-day spectrographs\cite{cote99a,cote99b}.  Large area surveys in search
of low surface brightness objects have led to the discovery of three new M31
dSph galaxies in the last couple of years\cite{armand98,kara,armand99},
bringing the census to six dSph galaxies.  The present set of M31 dSph
galaxies cover the same range of galactocentric distances as the Milky Way
companions.  Thus it should be possible to make a direct comparison of
environmental effects on these two sets of dSphs.

We present initial results from a Keck spectroscopic survey of five of the
six M31 dSph companions, And\,I, And\,III, And\,V, And\,VI, and And\,VII.
The first four have been studied with LRIS using a similar multislit Ca~II
triplet setup as for our M31 field halo star survey (Sec.~2.3); the fifth
dSph has been observed with the High-Resolution Echelle
Spectrometer\cite{vogt} (HIRES)---see Sec.~3.2.4 below.

\subsection{Preliminary Results}
\label{sect:3.2}

\subsubsection{Integrated radial velocity}
\label{sect:3.2.1}

Integrated stellar velocities of the five M31 dSph companions are listed in
Table~1.  In each case, this is based on the mean radial velocity of
15--25~member giants.  The quoted uncertainty is the $1\sigma$ error of the
mean, derived from the observed rms dispersion of velocities among member
stars (combination of measurement error and internal velocity dispersion); it
does {\em not\/} include possible systematic error in the measurements.
Note, the signal-to-noise ratio of the And\,III data is substantially lower
than that of the other dSphs and this is due to truncation of the exposure
time.

\begin{table} [h]   
\caption{Mean heliocentric velocities for five M31 dwarf spheroidal galaxies}
\label{tab:vels}
\begin{center}       
\begin{tabular}{|l|l|} 
\hline
\rule[-1ex]{0pt}{3.5ex}  Name & $v_{\rm hel}$ (km~s$^{-1}$)     \\
\hline
\rule[-1ex]{0pt}{3.5ex}  And\,I               & $-369.5\pm2.2$  \\
\rule[-1ex]{0pt}{3.5ex}  And\,III             & $-352.3\pm13.6$ \\
\rule[-1ex]{0pt}{3.5ex}  And\,V               & $-387.0\pm4.0$  \\
\rule[-1ex]{0pt}{3.5ex}  And\,VI (Peg\,dSph)  & $-340.7\pm2.9$  \\
\rule[-1ex]{0pt}{3.5ex}  And\,VII (Cas\,dSph) & $-306.7\pm2.3$  \\
\hline 
\end{tabular}
\end{center}
\end{table}

It is interesting to compare the velocity measurements in Table~1 with a
recent neutral hydrogen survey described in Ref.~\citenum{blitz}.  The HI
cloud tentatively detected in the vicinity of And\,III has a radial velocity
of $v_{\rm LSR}=-337\pm6$~km~s$^{-1}$ in the radio 21 cm line, which is
within $1\sigma$ of the optical velocity measurement.  If this low
significance HI detection at And\,III is confirmed, it likely represents gas
that is physically associated with the dSph.  By contrast, the 21 cm velocity
of the HVC~368 cloud in the direction of And\,V is $v_{\rm LSR}=-176\pm1$ and
this is clearly inconsistent with the optical value.  As originally
suspected\cite{blitz}, this cloud is likely to be a fragment of a larger
neighboring gas cloud (complex `H') that happens to be superposed on And\,V.

\subsubsection{Metallicity}
\label{sect:3.2.2}

Figure~5 shows CMDs of four of the M31 dSphs, the ones we have targeted for
LRIS spectroscopy (targets marked by special symbols).  The slope and width
of the red giant branch traced by member stars are indicative of the mean
metallicity and metallicity spread of the dSph, respectively.  For example,
the And\,V red giant branch is significantly closer to vertical than And\,I's
or And\,VI's indicating a lower metallicity.  This is confirmed by the
spectroscopic [Fe/H] estimates\cite{ggp} for these galaxies (see Fig.~6).

\begin{figure}
\plotfiddle{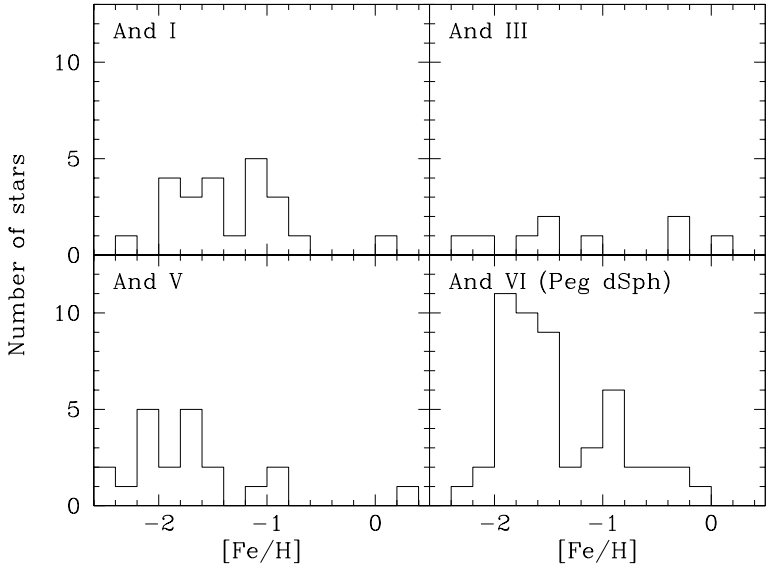}{1.7truein}{-90}{40}{40}{-160}{153}
\vskip 2.5truecm
\caption[m31dsph_feh]{ \label{fig:m31dsph_feh}
Distribution of [Fe/H] among the member red giants in four M31 dSph
companions from Keck/LRIS observations\cite{ggp}.  These [Fe/H] estimates are
based on the Ca~II absorption line strength calibrated via the CG97
formula\cite{carretta} given in Ref.~\citenum{rutledge97b}.  The And\,VI
spectra have the highest signal-to-noise ratio of the four, while the
And\,III spectra have the lowest signal-to-noise.}
\end{figure}

Perhaps an even more striking aspect of Fig.~5 is the variation of the {\em
width\/} of the red giant branch from one dSph to another.  It is convenient
to compare And\,V to And\,VI since their CMDs are based on
photometric data of near-identical quality: short $V$ and $I$ band exposures
with LRIS in $0.6''$--$0.8''$ seeing conditions\cite{gg98,gg99}.  While the
And\,V member giants define a very narrow track, the And\,VI members display
a substantial spread in color that is significantly in excess of the
photometric error; this is most easily explained in terms of a metallicity
spread in And\,VI.  The spectroscopically-determined [Fe/H] histograms
(Fig.~6) support this hypothesis: despite the fact that it has the highest
signal-to-noise spectra of the four, the And\,VI histogram shows a prominent
tail towards high metallicity and a spread in excess of 1.5~dex.  Similarly,
the And\,I red giant branch is broader than that of And\,III (Fig.~5),
perhaps due to an intrinsic metallicity spread in the former galaxy.  The
large spread in spectroscopic [Fe/H] values observed for And\,III (Fig.~6) is
likely a result of measurement error caused by the low signal-to-noise ratio
of its spectra.

\subsubsection{Dynamical mass estimate for the extended M31 halo}
\label{sect:3.2.3}

Dwarf companions can be treated as test particles in a large galaxy's
gravitational potential.  Ref.~\citenum{evans} presents the most up-to-date
and thorough dynamical analysis of the M31 halo using available radial
velocity and distance data for several classes of tracers: gas and stars in
the rotating disk, globular clusters, planetary nebulae, and dwarf
satellites.  The recent discovery of distant dSphs around M31, some almost
300~kpc from the parent galaxy, has breathed new life into this subject.  The
most distant tracers provide the best constraints on the total mass of the
halo.  An update of the M31 dynamical analysis, including radial velocity
data for all of the dSphs presented in this paper\cite{ewggv}, yields the
best total mass estimate to date: $\sim8\times10^{11}\,M_\odot$.
Surprisingly, this M31 value is somewhat lower than the best estimate for the
total mass of the Milky Way halo.  However, the large uncertainty associated
with each determination means that one cannot rule out the possibility that
the two galaxies have the same mass.

\subsubsection{Internal dynamics of And\,VII: Multislit HIRES experiment}
\label{sect:3.2.4}

The HIRES observations of And\,VII red giants presented in this paper were
carried out in an unusual manner. Special purpose multislit masks were
designed and fabricated to enable simultaneous observations of multiple red
giants.  Each mask contains 5--6 slitlets, with lengths in the range $1''$ to
$2.5''$, so that the sum total of all slitlet lengths is $\leq11''$ in order
to prevent overlap of adjacent echelle orders.  Needless to say, there is not
much blank ``sky'' area available on the typical slitlet.  Instead of the
usual method of sky subtraction, we synthesize a single sky spectrum for each
multislit mask by coadding blank pixels from all the slitlets on that mask.
The experiment was made doubly difficult by the fact that our typical target
star is very faint by HIRES standards: $V\sim22$--23.  Nevertheless, it is
possible to reliably extract velocities for most of the 21~stars (Fig.~7).
The observed rms dispersion is $\sigma_v=9.4$~km~s$^{-1}$, quadrature sum of
the dSph's internal velocity dispersion and the velocity measurement error.
These data are being used to probe the dynamics and mass-to-light ratio of
And\,VII\cite{ggvew}.

\begin{figure}
\plotfiddle{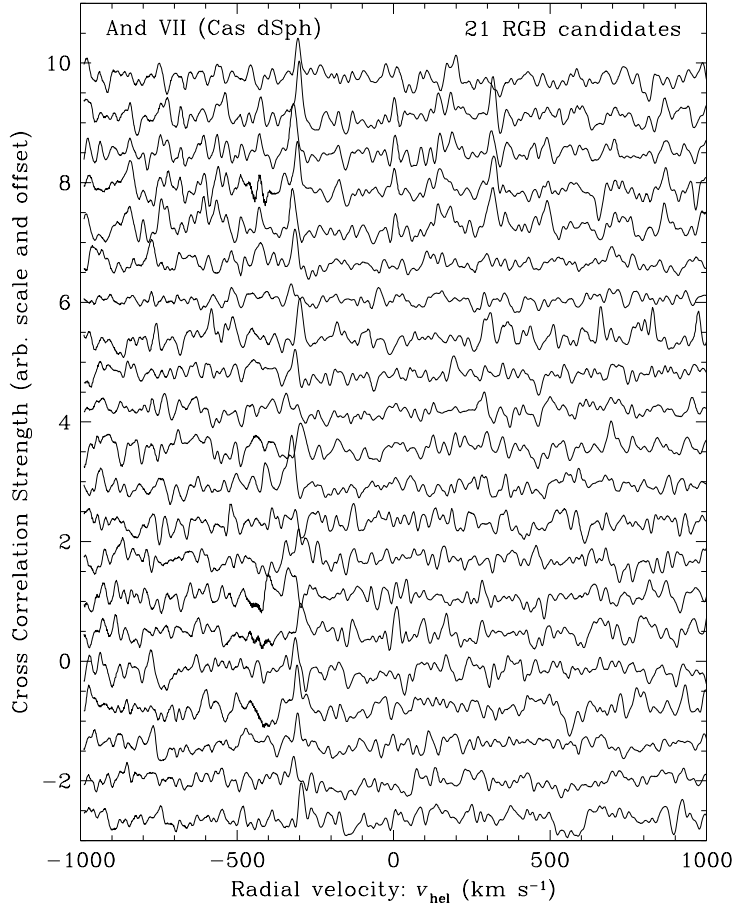}{2.447truein}{0}{49}{49}{-150}{-175}
\vskip 5.0truecm
\caption[and7_ccf]{ \label{fig:and7_ccf}
Cross-correlation functions for 21 red giant candidates in And\,VII derived
from HIRES multislit echelle observations.  The peaks are concentrated in the
vicinity of $v\approx-300$~km~s$^{-1}$, marking the systemic
velocity of the dSph.  Radial velocity variations from star to
star are due to the galaxy's internal velocity dispersion.}
\end{figure}

\section{SUMMARY}
\label{sect:summary}

\subsection{Main Points: M31 Halo Red Giants}
\label{sect:4.1}

The first part of this paper describes a study of field red giant stars in
the outer halo of M31:\cite{reitzel98,reitzel00}

\begin{itemize}

\item[$\bullet$]
A new color-based screening technique, using deep $UBRI$ photometry obtained
with the KPNO 4-meter telescope, is combined with traditional morphological
star-galaxy separation based on high resolution $I$-band images obtained
with the Keck 10-meter/LRIS.  This enables us to isolate M31 halo red giant
stars in a field located at a projected distance of 19~kpc from the galaxy's
center along the minor axis.  After screening for stars, this field displays
a clear excess population relative to a well-matched comparison field,
plausibly red giants at the distance of M31.

\item[$\bullet$]
These data, taken together with {\it Hubble Space Telescope\/} studies of the
inner halo, suggest that the density of M31's stellar halo is about
$10\times$ higher than that of the Milky Way halo at comparable radii.

\item[$\bullet$]
Follow-up Keck LRIS spectra are presented for 99~M31 halo red giant
candidates and a control sample of 12~foreground Galactic dwarf stars.
Kinematical information derived from the spectra allow elimination of
residual contaminants: foreground Galactic dwarfs, M31 disk giants,
background galaxies.

\item[$\bullet$]
Metallicity measurements are made on a star-by-star basis using two
independent methods: the strength of the Ca~II absorption lines and the
location in the $B-I$ vs $I$ color-magnitude diagram.  Various calibration
schemes are tried in deriving spectroscopic [Fe/H] estimates.  The
photometric and spectroscopic [Fe/H] estimates are in reasonable agreement
with each other.

\item[$\bullet$]
The mean metallicity of M31 field halo giants, $\rm\langle[Fe/H]\rangle$, is
in the range $-1.6$ to $-1.3$ depending on the metallicity measurement
method, in rough agreement with the mean value of the Milky Way halo.  There
does not appear to be a strong radial abundance gradient in M31's halo: the
mean [Fe/H] in this $R=19$~kpc halo field is similar to (possibly slightly
lower than) {\it HST\/} findings for the inner parts of the halo
($R=7$--11\,kpc).

\item[$\bullet$]
There is a spread of at least 2~dex in [Fe/H] even for the
spectroscopically-selected sample of secure M31 halo red giants, independent
of the method used to measure [Fe/H].  This is comparable to the [Fe/H]
spread among M31 globular clusters and Local Group dwarf satellite galaxies.

\item[$\bullet$]
The velocity distribution in the $R=19$~kpc minor axis halo field is well fit
by an equal mix of foreground dwarfs (drawn from a standard Galactic model)
and giants in M31's halo represented by a Gaussian of width
$\sigma_v\sim150$~km~s$^{-1}$ centered on its systemic velocity ($v_{\rm
sys}=-297$~km~s$^{-1}$).

\end{itemize}

\subsection{Main Points: M31's Dwarf Spheroidal Satellites}
\label{sect:4.2}

The second part of the paper presents initial results from an ongoing
Keck survey of red giants in dwarf spheroidal satellites of M31, including
three newly discovered members:\cite{ggprpvom,ewggv,ggvew,ggp}

\begin{itemize}

\item[$\bullet$]
We present spectra of about 100~red giant candidates in five M31 dSph
companions: And\,I, And\,III, And\,V, And\,VI (Peg\,dSph), and And\,VII
(Cas\,dSph).  The first four have been observed with LRIS and the last one
with HIRES.  This fall, we plan to carry out observations with ESI, the new,
intermediate-resolution, high-throughput spectrograph on Keck.

\item[$\bullet$]
The mean stellar radial velocities of the dSphs are compared to recent
velocity measurements of neutral hydrogen clouds seen in the direction of
some of these dwarfs\cite{blitz}.  The stellar and HI velocities are in good
agreement in the case of And\,III, indicating a physical association, but not
in the case of And\,V (probably a chance superposition).

\item[$\bullet$]
Color-magnitude diagrams of spectroscopically-confirmed dSph member giants
suggests the presence of a significant metallicity spread in And\,VI and
possibly in And\,I, but little or no spread in And\,III and And\,V.
Spectroscopic estimates of the metallicity distribution in these four dSph
galaxies, based on the strength of the Ca~II triplet in member red giant
stars, appear to be consistent with this finding.

\item[$\bullet$]
The And\,VII observations are based on a novel multislit (5--6 stars at a
time), echelle spectroscopy technique using HIRES.  Accurate radial velocity
measurements have been obtained for a total of 15--20~member giants.  The
observed rms velocity dispersion is about $\sigma_v\sim9$~km~s$^{-1}$
(uncorrected for measurement error).

\item[$\bullet$]
Using our new radial velocity measurements of these outer dSph satellites in
combination with other dynamical tracers, the total mass of M31's extended
halo is estimated to be a little under $10^{12}\,M_\odot$, comparable to or
smaller than that of the Milky Way halo.

\end{itemize}

\acknowledgments     
 
P.G.\ would like to thank his collaborators on various ongoing projects from
which the data and results in this paper are drawn: Steve Majewski, Jamie
Ostheimer, Drew Phillips, Mark Wilkinson, and especially Wyn Evans, Linda
Pittroff, and Steve Vogt.  He is grateful to the organizers of the SPIE
meeting in Munich for their generous hospitality.  E.K.G.\ acknowledges
support by NASA through grant HF-01108.01-98A from the Space Telescope
Science Institute.



  \end{document}